\documentclass[referee]{raa} 
\usepackage{natbib}
\usepackage{graphicx,times}   
\usepackage{amssymb,amsmath}
\usepackage{threeparttable}
\usepackage[pass]{geometry}
\usepackage{pdflscape}
\usepackage{multirow}
\usepackage[colorlinks=true, linkcolor=blue, citecolor=blue, urlcolor=blue]{hyperref}
\usepackage{float}
\usepackage{CJK}

\begin{document}
\begin{CJK*}{UTF8}{gbsn}

   \title{Limiting Magnitudes of the Wide Field Survey Telescope (WFST)
%\footnotetext{$*$ Supported by the National Natural Science Foundation of China.}
}
%   \subtitle{I. Place Your Subtitle Here}

   \volnopage{Vol.0 (21xx) No.0, 000--000}      %%preserved for Editor. DOn't remove!
   \setcounter{page}{1}          %%starting page, preserved for Editor. DOn't remove!
   \author{Lei Lei (雷磊)\inst{1,2} % （雷磊） astroLL@mail.ustc.edu.cn
    \and Qing-Feng Zhu (朱青峰)\inst{1,3} % （朱青峰） zhuqf@ustc.edu.cn
    \and Xu Kong (孔旭)\inst{1,3} % （孔旭） xkong@ustc.edu.cn
    \and Ting-Gui Wang (王挺贵)\inst{1,3} % （王挺贵） twang@ustc.edu.cn
    \and Xian-Zhong Zheng (郑宪忠)\inst{1,2} % （郑宪忠） xzzheng@pmo.ac.cn
    \and Dong-Dong Shi (师冬冬)\inst{2} % （师冬冬） ddshi@pmo.ac.cn
    \and Lu-Lu Fan (范璐璐)\inst{1,3} % （范璐璐） llfan@ustc.edu.cn
    \and Wei Liu (刘伟)\inst{2} % （刘伟） liuwei@pmo.ac.cn
   }

   \institute{School of Astronomy and Space Science, University of Science and Technology of China, Hefei 230026, China; {\it zhuqf@ustc.edu.cn}\\
        \and
        Purple Mountain Observatory, Chinese Academy of Sciences, Nanjing 210023, China;\\
        \and 
        Deep Space Exploration Laboratory / Department of Astronomy, University of Science and Technology of China, Hefei 230026, China;\\
%%        \and
%%             Full institute address for the third author\\
\vs\no
   {\small Received~~2022 September 19; accepted~~2022~~December 23}}

\abstract{ Expected to be of the highest survey power telescope in the northern hemisphere, the Wide Field Survey Telescope (WFST) will begin its routine observations of the northern sky since 2023. WFST will produce a lot of scientific data to support the researches of time-domain astronomy, asteroids and the solar system, galaxy formation and cosmology and so on. We estimated that the 5 $\sigma$ limiting magnitudes of WFST with 30 second exposure are 
$u=22.31$ mag, $g=23.42$ mag, $r=22.95$ mag, $i=22.43$ mag, $z=21.50$ mag, $w=23.61$ mag. The above values are calculated for the conditions of $airmass=1.2$, seeing = 0.75 arcsec, precipitable water vapour
(PWV) = 2.5 mm and Moon-object separation = $45^{\circ}$ at the darkest New Moon night of the Lenghu site (V=22.30 mag, Moon phase $\theta=0^{\circ}$). The limiting magnitudes in different Moon phase conditions are also calculated. The calculations are based on the empirical transmittance data of WFST optics, the vendor provided CCD quantum efficiency, the atmospherical model transmittance and spectrum of the site. In the absence of measurement data such as sky transmittance and spectrum, we use model data.  %Further, the limiting magnitudes results show that the limiting magnitudes of WFST change very little with seeing varies from 0.3 to 2.0 arcsec, and the WFST maintain a high limiting sensitivity, which is beneficial to the survey in the future. 
%A site condition monitoring program showed that the night sky brightness can reach V=$22.3$ $\rm mag/arcsec^2$ during a fully clear new moon time in high-latitude area, and the Lenghu site is among the first class optical astronomical sites in the world. Based on the optical design of WFST, the atmospheric transmission, the vendor provided CCD quantum efficiency, and the estimate of the atmospherical spectrum of the site, we compute the limiting magnitudes of WFST.
}
   \authorrunning{Lei et al.}%The WFST Collaboration            %author_head in even pages
   \titlerunning{Limiting Magnitudes of WFST}  % title_head in odd pages
   \maketitle
%% The author head (on even pages) and the title head (on odd pages) will be
%% automatically extracted from \author{} and \title{}. Whenever the title is too long,
%% you will be asked to supply a shorter one by inserting either \authorrunning{} or
%% \titlerunning{} before \maketitle. Anyway, you can specify your own heads.
%
%% Note: In the following text body of your manuscript, please note several differences fromSesar+etal+2010
%%       other major journals:
%% (1) \subsection{Please Capitalize the First Letter of Each Notional Word in Subsection Title}
%% (2) Please Capitalize the First Letter of Each Notional Word in all tables' captions
%
%________________________________________________ sections below
%
\section{Introduction}           %% first-level sections will be auto-capitalized
\label{sect:intro}
The Wide Field Survey Telescope (WFST; \citealt{Lou+etal+2016}; \citealt{Shi+etal+2018}; \citealt{Lou+etal+2020}; \citealt{Lin+etal+2022}) is an optical telescope to be installed at the Lenghu site, located on Saishiteng mountain near Lenghu Town in Qinghai Province on the Tibetan Plateau, China in 2023. The WFST has a 2.5-m diameter primary mirror and a 5-lens corrector to form a prime-focus optics (\citealt{Wang+etal+2016};\citealt{Lou+etal+2016}; \citealt{Lou+etal+2020}). The detector of WFST consists of nine 9K$\times$9K CCD chips and has 0.9 Giga pixels. The entire system is optimized for the wavelength range from 3200 $\textsc{\AA}$ to 9600 $\textsc{\AA}$ (\citealt{Lou+etal+2016}; \citealt{Chen+etal+2019}). With the aid of an active optics system and an ADC (atmospheric dispersion compensator), WFST can achieve an image quality of 0.4 arcsec 80\% energy enclosed across a field of view of 3 degree diameter and $\sim$7 deg$^2$ area.
WFST is able to survey $\sim2\times10^4$ deg$^2$ northern sky in $ugrizw$ six bands. 
As a powerful survey telescope, its scientific data will greatly support researches of time-domain astronomy, asteroids and the solar system, the Milky Way and its satellite dwarf galaxies, galaxy formation and cosmology and so on.%The camera is made a mosaic charge-coupled device (CCD) detector using nine CCDs and each has 9k {$\times$} 9k pixels.

In recent years, many large ground-based optical survey telescopes have been built or planned all over the world. SDSS (\citealt{Kent+etal+1994}; \citealt{Fukugita+etal+1996}), Pan-STARRS (\citealt{Jedicke+etal+2007}; \citealt{Chambers+etal+2016}), SkyMapper (\citealt{Schmidt+etal+2005}; \citealt{Rakich+etal+2006}), ZTF (\citealt{Bellm+etal+2014}; \citealt{Bellm+etal+2019}; \citealt{Graham+etal+2019}) and other built telescopes have produced a large amount of observation data, which has greatly promoted astronomical researches and solved many scientific problems. Soon new, survey telescopes such as LSST (\citealt{Hlozek+etal+2019}), Mephisto (\citealt{Liu+etal+2019}; \citealt{Lei+etal+2020}; \citealt{Yuan+etal+2020}) and WFST will join their peers and conduct deeper multi-band surveys to provide crucial data to astrophysical researches. Combined with China Space Station Telescope (CSST; \citealt{Zhao+etal+2016}; \citealt{Yuan+etal+2021}) and other space telescopes, WFST will greatly improve human understanding of the universe and promote more important scientific discoveries. %the Nancy Grace Roman Space Telescope (\citealt{McEnery+etal+2021}) %China has begin to build 8 meter solar telescope (\citealt{Liu+etal+2012}; \citealt{Deng+etal+2016}) and Chinese 12m optical/infrared telescope (LOT, \citealt{Cui+etal+2018}). To estimate of sensitivity for large infrared telescopes based on measured sky brightness and atmospheric extinction, \citet{Zhao+etal+2021} calculated limiting magnitudes of 10m telescopes without instrumental emission. Their result is 13.01 mag at Ali and 12.96 mag at Daocheng.

The parameters of a survey telescope, such as the diameter of the primary mirror, quantum efficiency (QE) of CCD, band transmittance, etc., determine the throughput of the telescope. The site conditions of an astronomical observatory, such as atmospheric transmittance, altitude, seeing and sky background brightness, affect the depth of a survey program.. The limiting magnitude of a survey telescope is an important guide for planning research objectives and project scopes. It is also the key for designing exposure time plans and survey strategies. Therefore, an accurate estimation of the limiting magnitudes are needed for the successful commission of a new telescope.

In this work, we introduce the estimation of the limiting magnitudes of WFST. In Sect.~\ref{sect:LM} we describe the method we adopt.
In Sect.~\ref{sect:conclusion} we show our results of limiting magnitudes of the WFST.

\section{Limiting magnitudes}
\label{sect:LM}
%The limiting magnitude of an astronomical telescope is mainly determined by the image quality, the sky background noise, the area of the primary mirror and the throughput of the system. %quantum efficiency of the detector. 
\subsection{Throughput of WFST}
The throughput of an astronomical observation ($T_{tot}$) is limited by the atmospheric transmittance ($T_{atmo}$), the transmittance of optics ($T_{opt}$), the transmittance of the filters $T_{band}$) and the quantum efficiency of CCD ($QE_{CCD}$). 
\begin{equation}
\label{eq_throughput}
    T_{tot}=T_{atmo}\cdot T_{opt}\cdot T_{band}\cdot QE_{CCD}
\end{equation}

The optical system of WFST consists of a 2.5 meter diameter primary mirror with a 760 mm diameter central hole, five corrector lenses, a ADC made with two glass wedges and $ugrizw$ six switchable filters \citep{Lou+etal+2016}. Among five correcting lenses, only one is made of the N-BK7HT glass. The others and ADCs are made of the fused quartz. Since the transmittance of a fused quartz blank can be neglected, we simulate the total transmittance of the five-lens corrector and the ADC with the product of the transmittance of a 35 mm thick N-BK7 glass blank and the transmittance of 14 layers of anti-reflection (AR) coatings. The transmittance of the N-BK7 glass is obtained from \emph{SCHOTT}$\footnote{\url{https://refractiveindex.info/?shelf=glass&book=BK7&page=SCHOTT}}\label{footnote0}$ and the transmittance of the AR coating is from Institute of Optics and Electronics (IOE) 's measurements. Because of the oversized Primary Focus Assembly (PFA), the actual aperture obscuration is 1 m diameter.

\begin{figure}[h]
\centering
\includegraphics[width=0.65\textwidth, angle=0]{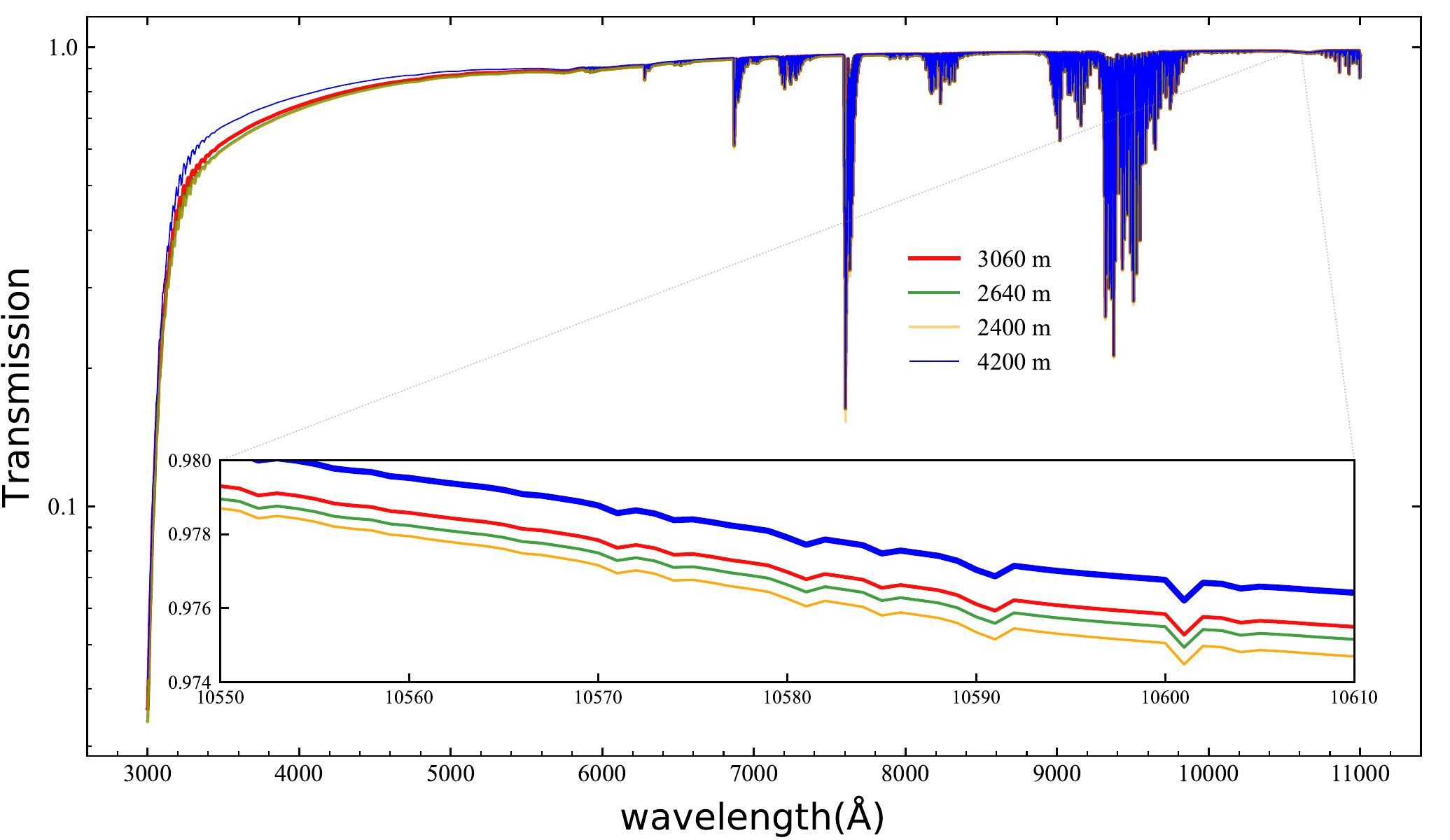}
\caption{The atmospheric transmittance curves of \emph{Paranal} Observatory (2400 m), \emph{La Silla} Observatory (2600 m) and the \emph{Extremely Large Telescope} site (3060 m), the scaled transmittance curve of the WFST site (4200 m). We assume the WFST site has the same geographic features (i.e. high altitude Mountain and dry air) as the three ESO sites.}
\label{Fig1}
\end{figure}

Because we don't have atmospheric transmitance and spectrum measurements at the site, we adopt $\emph{SkyCalc}\footnote{\url{https://www.eso.org/observing/etc/bin/gen/form?INS.MODE=swspectr+INS.NAME=SKYCALC}}$ (Version 2.0.9) to obtain model curves. \emph{SkyCalc} is developed by astronomers in ESO based on the \emph{Cerro Paranal Advanced Sky Model} (\citealt{Noll+etal+2012}; \citealt{Jones+etal+2013}; \citealt{Moehler+etal+2014}). %\label{footnote1}
The atmospheric transmittance is affected by altitude, humidity, dust, precipitable water vapour (PWV), among which the altitude is the most important factor. \emph{SkyCalc} only provides the atmospheric transmittance at three astronomical sites: \emph{Paranal} (2400 m), \emph{La Silla} (2600 m) and \emph{Extremely Large Telescope (ELT)} site (3060 m). 
Figure \ref{Fig1} shows the three transmittance curves of the sites. We can see that three curves have same features that they are scaled according to different altitudes of three sites. This is reasonable because the geographic features of the three sites are very similar.
The \emph{Paranal} Observatory is on the Cerro Paranal mountain, which is in the Atacama Desert of northern Chile.
The \emph{La Silla} Observatory is located on the outskirts of the Chilean Atacama Desert.
The 40-metre-class \emph{ELT} is on the Cerro Armazones mountain in the central part of the Atacama Desert. WFST is on the Saishiteng mountain in the Gobi desert area on the Tibetan Plateau. We consider that the geographic features of the area are more similar to those of sites in the high-altitude Atacama desert than those of oceanic mountain areas, such as Mauna Kea in Hawaii. It is a reasonable choice to obtain the atmospheric transmittance of the WFST site by using the spectra from \emph{SkyCalc}.
So we get the atmospheric transmittance curve of the WFST site at an altitude of 4200 m by scaling the atmospheric transmittance curves of \emph{paranal}, \emph{lasilla} and ELT sites.
In our simulations, we assume that airmass = 1.0 and precipitable water vapour (PWV) = 2.5 $\rm mm$. Figure \ref{Fig1} also shows the result of scaling. 

\begin{figure}[h]  %[htbp]中的h是浮动的意思
    \centering    %居中
    \includegraphics[width=0.49\textwidth]{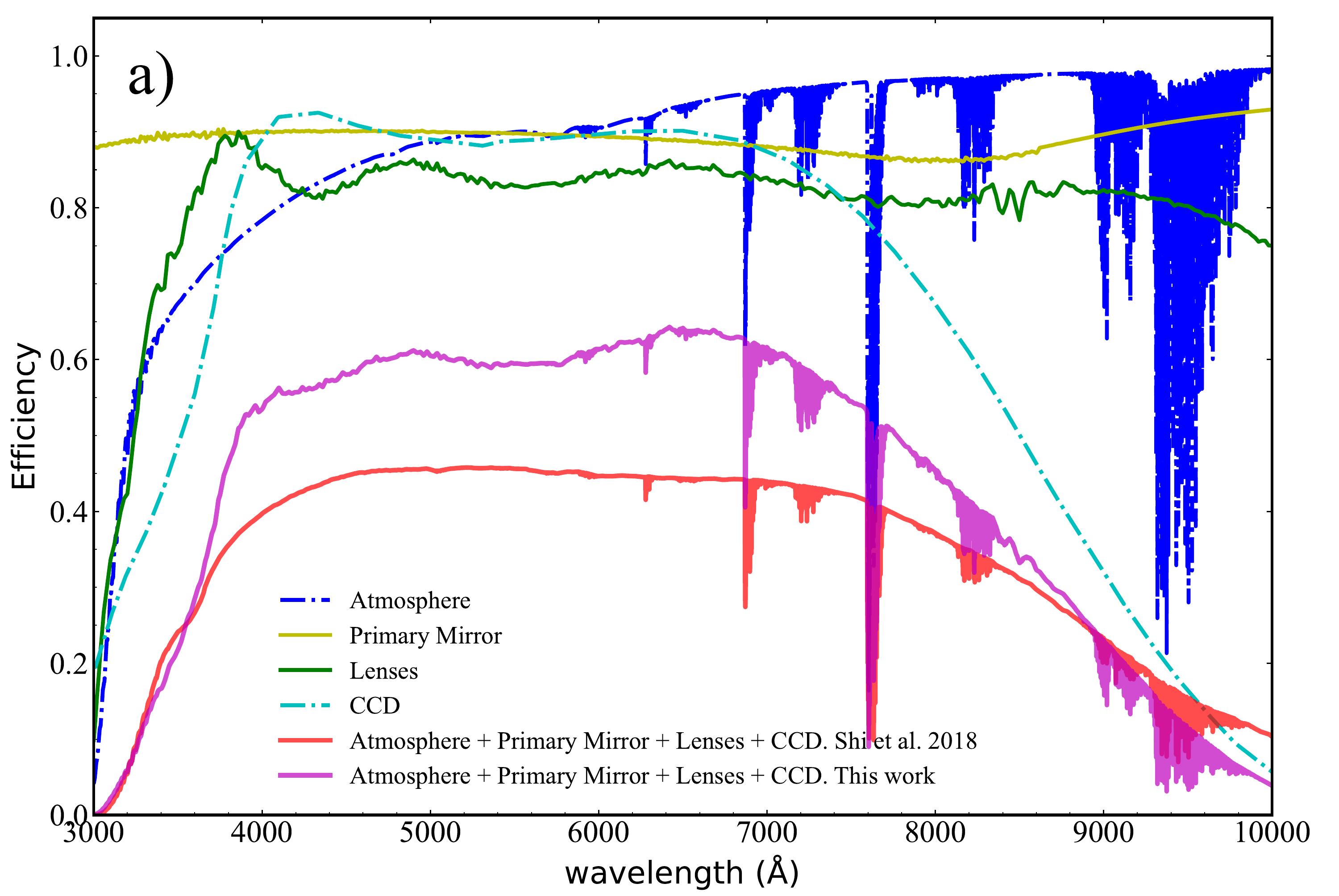}   %以行宽的0.5倍大小显示
    \includegraphics[width=0.49\textwidth]{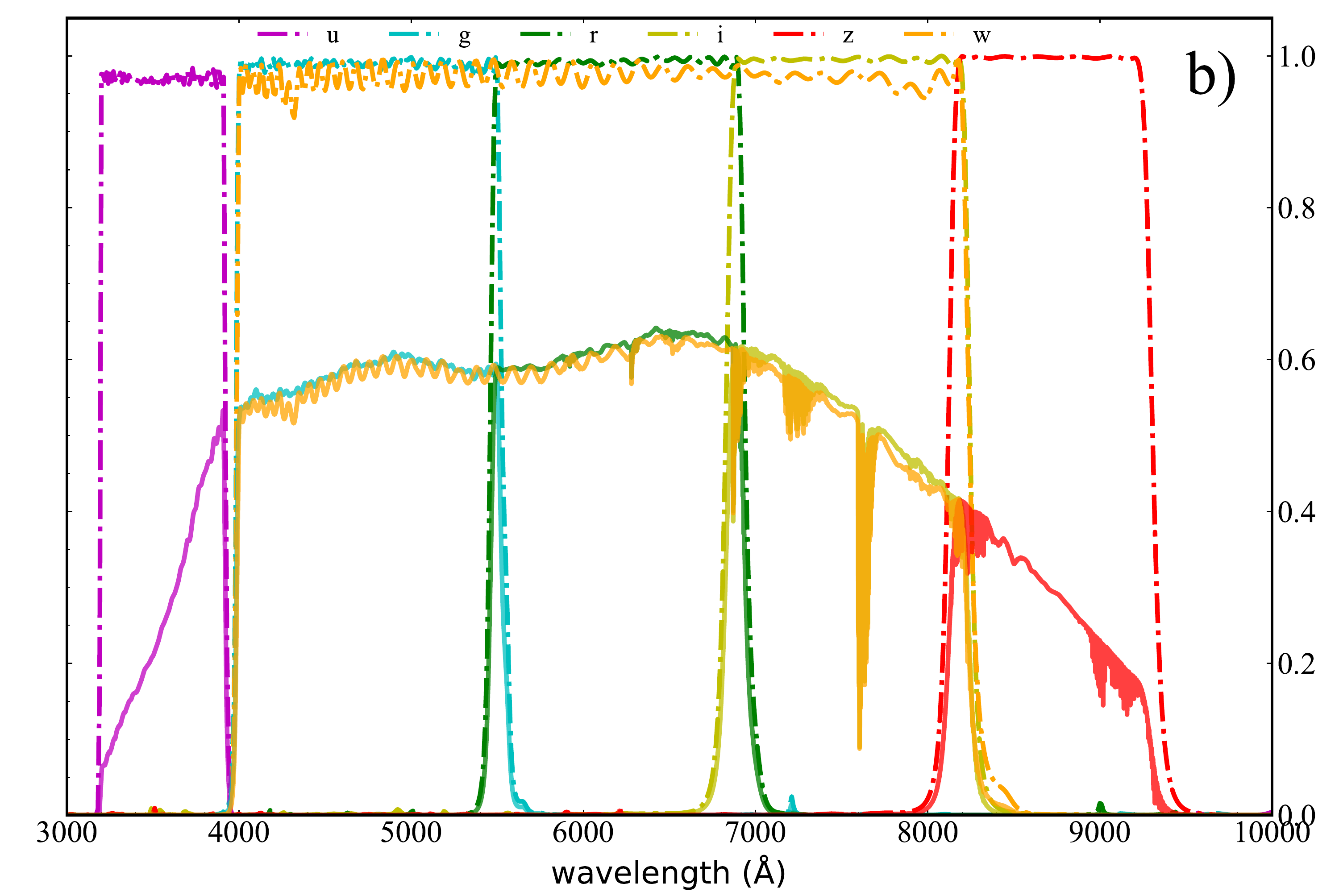}   %以行宽的0.5倍大小显示%
    \caption{(a) The transmittance curves of the atmosphere (blue dot-dashed line), the optics including lenses and the ADC (green line), the reflectivity of the primary mirror (yellow line) and the quantum efficiency of the CCDs (cyan dot-dashed line). The combined efficiency of the atmosphere, the optics and the CCDs is in purple. The red line shows the total efficiency of Shi et al.(2018); (b) The transmittance of each filter and the total efficiency in each WFST filter transmittance.} %  %大图名称
    \label{Fig2}  %图片引用标记
\end{figure}

As shown in Figure \ref{Fig2}(a), combined system throughput, individual transmittance curves of the atmosphere and the corrector, the reflectivity of the primary mirror and the quantum efficiency of the CCD are plotted respectively. We also plot the original estimate of the system throughput for WFST by \citet{Shi+etal+2018}. We can see that the updated system throughput is higher than the early expectation in most wavelengths \citep{Shi+etal+2018}. The major reason is that the transmittances of ADC and optical lenses are higher than the early estimate. In order to obtain high efficiency in short wavelengths, WFST selects \emph{e2v} standard Si back-illuminated CCD detectors with the astro multi-2 coating. The QE of the CCDs is also increased. 

Figure \ref{Fig2}(b) shows the transmittance of the filters and the total throughput of WFST in six bands, which is calculated by using Equation \ref{eq_throughput}.

\subsection{Noise of WFST}

Noise in astronomical CCD images mainly consists of the contributions from the artificial light on the ground, astrophysical sources, sky background, CCD dark current, and CCD readout noise.
The site of WFST is on Saishiteng mountain on the Tibetan Plateau, where the nearest residential area is Lenghu town which is $\sim$ 50 km from the observatory site and has a population of 200. There is no industrial activity and the ground light pollution. The Haixi Mongolian and Tibetan Autonomous Prefecture of Qinghai Province has announced the 17800 square kilometres area of Lenghu as a dark night protecting region in the local law. It protects the good observational conditions of the Lenghu astronomical site. \citet{Deng+etal+2021} studied long-term astronomical conditions of the Lenghu site and pointed out that the sky background of a New Moon night can reach 22.3 $\rm mag\, arcsec^{-2}$ in the V band and the average night-sky brightness is around 22.0 $\rm mag\, arcsec^{-2}$ when the Moon is below the horizon. We adopt AB magnitude system in this work.

\begin{figure}[htb]  %[htbp]中的h是浮动的意思
    \centering    %居中
    \includegraphics[width=0.49\textwidth]{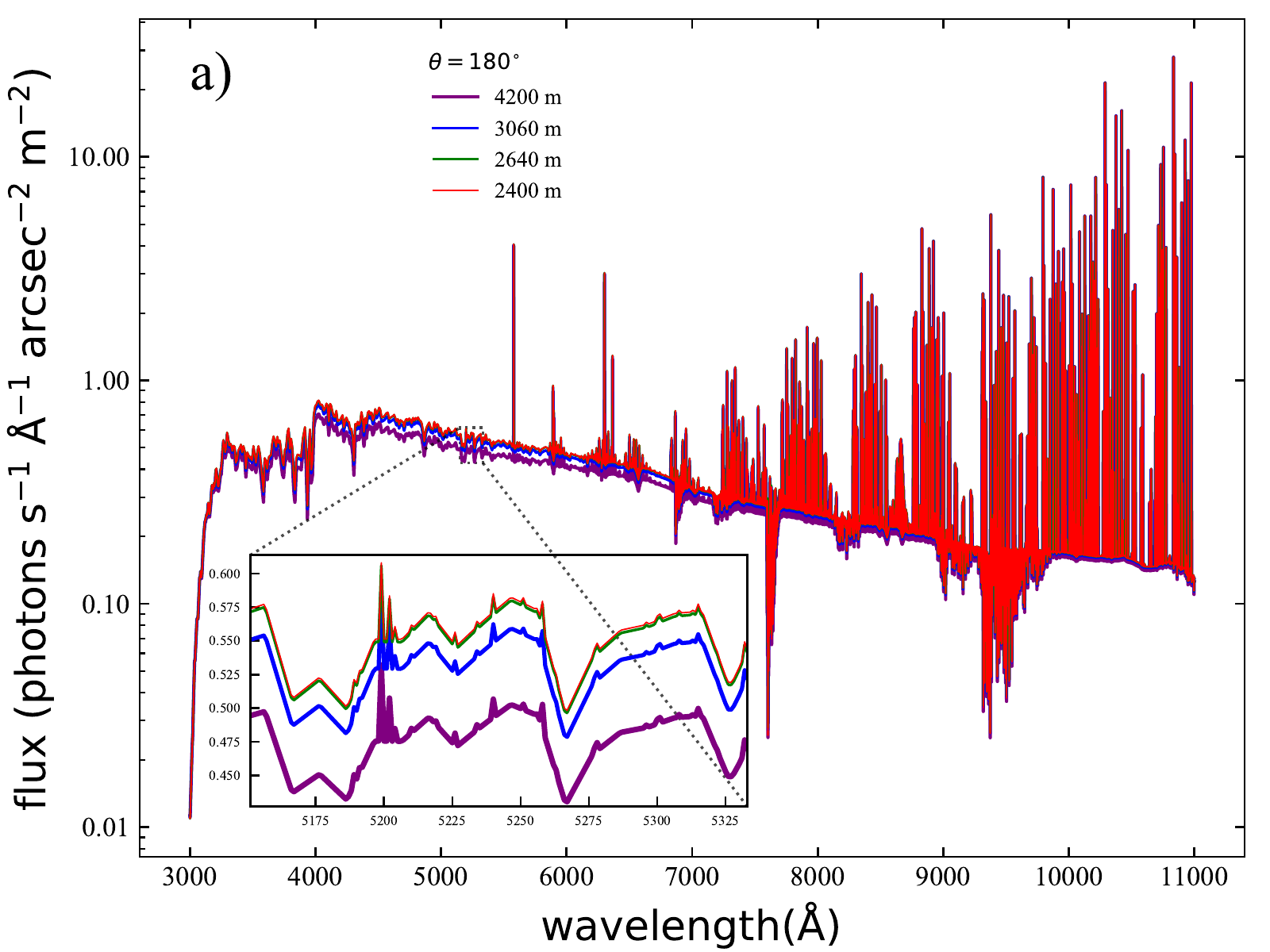}   
    \includegraphics[width=0.49\textwidth]{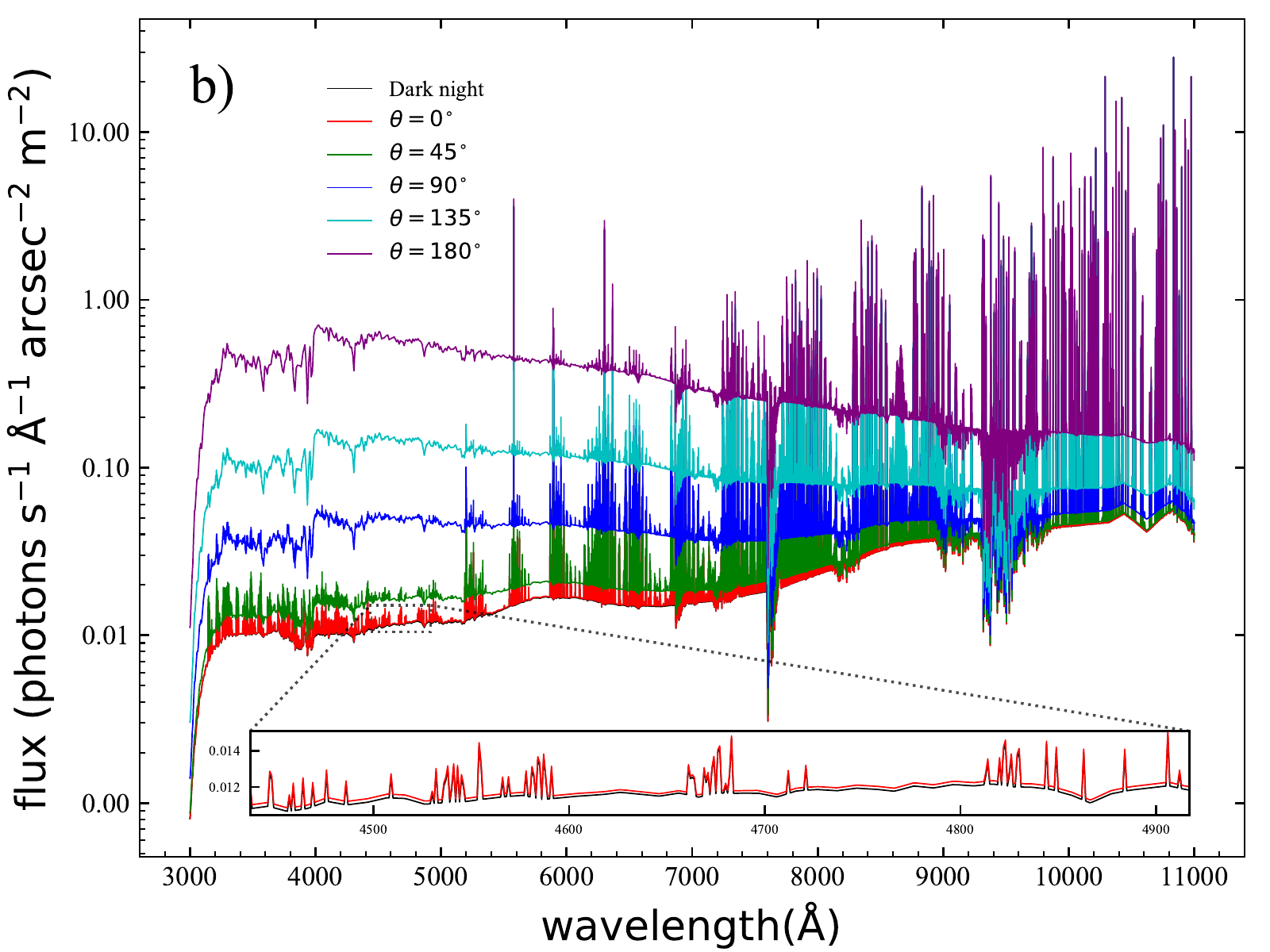}
    \caption{(a) The sky background spectrum (purple) at Zenith at the altitude of 4200 m in the Full Moon condition, and the spectra of three ESO sites when the Moon phase is $\theta=180^{\circ}$.
    (b) The Zenith sky spectra of the 4200 m site in different Moon phase conditions. The sky background spectrum of Moon phase $\theta=180^{\circ}$ and the spectrum of a dark night (when the Moon is under the horizon) have almost the same flux.} %  %大图名称
    \label{Fig3}  %图片引用标记
\end{figure}

%In cloudless and moonless nights, due to the excellent environment of Lenghu site, the observational noise of WFST is dominated by the sky background emission. In this case, combined with the telescope instrument parameters, we can calculate the observation noise using the spectrum of the sky background.  \citet{Jones+etal+2014} has simulated a sky spectrum template for Gemini telescope installed at Mauna Kea Observatory (MKO)$\footnote{\url{https://www.gemini.edu/observing/telescopes-and-sites/sites\#OptSky}}$. The spectrum is the sky at Mauna Kea is fainter than 20.78 mag/arcsec$^2$ for 50\% of the time (\citealt{Krisciunas+etal+1987}; \citealt{Krisciunas+etal+1991}; \citealt{Krisciunas+1997}).
The sky background spectrum of the Lenghu site is also calculated by the software \emph{SkyCalc} (\citealt{Noll+etal+2012}; \citealt{Jones+etal+2013}; \citealt{Moehler+etal+2014}). 
The monthly averaged solar radio flux is equal to 130.00 \emph{sfu}, that is the solar 10.7 cm radio flux in the Sun median active level (\citealt{Sparavigna+2008+etal}; \citealt{Petrova+2021+etal}). Because the solar activities will affect the sky background brightness, it is necessary to take the solar activities into account when we estimate the sky background. We adopt the median values obtained from long-term solar monitoring programs as the baseline solar brightness.
The spectral flux of one sky region is related to the Moon-Target separation and the Moon phase. 
We designate the Moon phase with the Moon phase angle ($\theta$) 0$^{\circ}$ (New Moon), 45$^{\circ}$ (Waxing), 90$^{\circ}$ (Half Moon Waxing), 135$^{\circ}$ (Waxing), 180$^{\circ}$ (Full Moon) respectively. 
We assume the separation of the Moon and a target is always 45$^{\circ}$ in our calculation. So the spectral flux of one region is only dependent on the altitude and the Moon phase. 
We get the sky background spectra towards the Zenith under different Moon phase conditions at the Lenghu Observatory site by scaling the spectra of three ESO sites provided by \emph{SkyCalc}. Figure \ref{Fig3}(a) shows the sky background spectrum at the altitude of 4200 m ($\theta=180^{\circ}$, here we just plot the spectra at a full Moon night because it is easier to see their difference.), and the spectra of the three ESO sites at Full Moon night. Figure \ref{Fig3}(b) shows the sky spectra at Lenghu site under six different Moon phase conditions. As shown in the detail part of Figure \ref{Fig3}(b), the sky background spectrum at a New Moon night ($\theta=0^{\circ}$) and the sky spectrum at a Dark night have almost the same flux.

%When the observational conditions of the two sites are similar and the same sky background spectrum template is adopted, here is a relationship between magnitude $m_{V}$ of the V band and the sky background spectrum flux $f_{\lambda}$.The V-band magnitudes of the 4200 m sky background $m_{V}$ is not equal to 22.3 mag. 

We can get the magnitude $m_{V}$ by integrating the sky background spectrum multiplied by the V band filter transmission curve:
\begin{equation}
\label{eq1}
    m_{V}=-2.5\times\left(\log_{10}\frac{\int_{0}^{\infty}f_{\lambda}T_{band,\lambda}d\lambda}{\int_{0}^{\infty}T_{band,\lambda}d\lambda}\right)-21.1
\end{equation}
where $f_{\lambda}$ is sky background spectral flux, $T_{band}$ is the \emph{Johnson} V band transmission curve \citep{Bessell+etal+1990}, $ZP=-21.1$ is the zero point (\citealt{Bessell+etal+2012}). The modeled sky emission radiance flux from \emph{SkyCalc} is in units of $\rm photon/s/m^2/micron/arcsec^2$.
The \emph{Johnson} V band sky background magnitude of a New Moon night with the \emph{SkyCalc} model spectrum at an altitude of 4200 m is 21.74 $\rm mag\, arcsec^{-2}$. We scale the 4200 m sky background spectrum so that the resulting spectrum has a V-band magnitude of 22.3 or 22.0 $\rm mag\, arcsec^{-2}$,
corresponding to the best and the average sky brightness conditions at the Lenghu site. 
As shown in Figure \ref{Fig3}(b), there are differences among the sky spectra under different Moon phase conditions. 
We scale these sky spectra at different Moon phases use the the same scaling factor in the new moon case, where we scale the spectrum from $V^{\theta=0^{\circ}}=21.74$ to 22.30 $\rm mag\, arcsec^{-2}$, so that differences among spectra at different Moon phases are not changed. The estimated V band Zenith sky background magnitudes at the Lenghu site with different Moon phases are: $V^{\theta=0^{\circ}}$, $V^{\theta=45^{\circ}}$, $V^{\theta=90^{\circ}}$, $V^{\theta=135^{\circ}}$, $V^{\theta=180^{\circ}}$=22.30, 22.10, 21.29, 20.28, 18.90 $\rm mag\, arcsec^{-2}$.

%Because the sky background brightness of Dark night is almost same as New Moon night (Moon phase $\theta=0^{\circ}$), in the following, we will no longer distinguish the difference between the two situations. We only list the results of the New Moon night instead of the two.  We scale the sky background spectrum from $V^{\theta=0^{\circ}}$=21.74 to $V^{\theta=0^{\circ}}$=22.3 $\rm mag\, arcsec^{-2}$. 
%Here is the relationship between magnitude $m_{V}$ of the V band and the sky background spectrum flux $f_{\lambda}$.
%\begin{equation}
%    20.83-m_{V}=-2.5\times\log_{10}\frac{f_{\lambda}}{f_{Lenghu,\lambda}}
%\end{equation}
%So we can get the sky background spectrum $f_{Lenghu,\lambda}$ of Lenghu site. 
The Lenghu sky background spectrum is calculated for $airmass$ = 1.0. It can be scaled to another \emph{airmass} by multiplying a factor $a$ (\citealt{Krisciunas+etal+1991}). 
%\begin{equation}
%    f_{Lenghu,\lambda}=a\times f_{4200m,\lambda}\times %10^{(m_{4200m,V}-m_{Lenghu,V})/2.5}
%\end{equation}
\begin{equation}
    a=10^{-0.172\frac{\left(X-1\right)X}{2.5}}
\end{equation}
when $airmass=1.2$, $X$ is 
\begin{equation}
    X=\frac{1}{\sqrt{(1-0.96\times \sin{(\arccos{(\frac{1}{airmass}})})^2)}}\approx 1.18958
\end{equation}
%Figure \ref{Fig4} shows the scaled sky background spectrum of the Lenghu site using the above method when the Moon phase is $\theta=0^{\circ}$ and the scaled sky background spectrum in each filter. The sky background spectrums are scaled into the condition $airmass$ = 1.2. %The black curves are the sky background spectrum at Zenith (cyan point), and the cyan curve is the spectrum at airmass=1.2 (black line).

%\begin{figure}[htb]
%   \centering
%   \includegraphics[width=0.65\textwidth, angle=0]{RAA/fig/fig_Sky_Band.pdf}
%   \caption{The sky background spectrum of New Moon night (Moon phase $\theta=0^{\circ}$) of Lenghu site when $m_V=21.74$ $\rm mag\, arcsec^{-2}$ (black dots) and $m_V=22.30$ $\rm mag\, arcsec^{-2}$ (cyan dots). The coloured lines show the sky background spectrum of Lenghu site of New Moon night in $ugrizw$ filters when $m_V=22.30$ $\rm mag\, arcsec^{-2}$. The black line shows spectrum at the New Moon night when $m_V=21.74$ $\rm mag\, arcsec^{-2}$.}
%   \label{Fig4}
%   \end{figure}

Based on the sky background spectrum of the Lenghu site, we estimate the magnitudes of the sky background in each band $m^{AB}_{band}$:
\begin{equation}
    m^{AB}_{band}=-2.5\times \log_{10}\left(\frac{Sky_{band}}{ZP_{band}}\right)%, band = u, g, r, i, z, w
\end{equation}
where
\begin{equation}
    ZP_{band}=\int_{0}^{\infty}flux_{AB}T_{band,\lambda}d\lambda
\end{equation}
where $flux_{AB} = 3631 Jy$ for all frequencies, and $T_{band,\lambda}$ is the transmittance curve of a particular band.
\begin{equation}
\label{eq_sky}
    Sky_{band}=\int_{0}^{\infty}f_{\lambda}T_{band,\lambda}d\lambda
\end{equation}
where $f_{\lambda}$ is sky background spectral flux.
%The zero point $ZP_{filter}$  (in unit of $\rm photons\; cm^{-2}\; s^{-1}$) in WFST $ugrizw$ bands are: $ZP_u=1.09\times10^6$, $ZP_g=1.79\times10^6$, $ZP_r=1.32\times10^6$, $ZP_i=1.05\times10^6$, $ZP_z=7.58\times10^5$, $ZP_w=3.88\times10^6$.  
%ZP0        [10.94 17.91 13.19 10.45  7.58 38.84] E+5
%Thus, the sky background magnitudes in AB magnitude system are obtained.
The Table \ref{tab0} shows the sky background magnitudes $m^{AB}$ in WFST six bands.
\begin{table}[h]
 \centering 
 \caption{The sky background brightness $m^{AB}$ of WFST six bands in units of $\rm mag\ arcsec^{-2}$.}
 \label{tab0} 
 \begin{tabular}{c c c c c c c c c} 
 \hline 
 \hline 
 Moon Phases & $u$ & $g$ & $r$ & $i$ & $z$ & $w$ \\
\hline 
%No Moon (under horizon)  & 23.00 & 22.58 & 21.82 & 21.05 & 19.30 &  21.68\\
0$^{\circ}$  & 23.27 & 22.82 & 21.80 & 20.99 & 20.05 & 21.78\\
45$^{\circ}$  & 23.02 & 22.49 & 21.66 & 20.93 & 20.03 & 21.64\\
90$^{\circ}$  & 22.00 & 21.37 & 20.99 & 20.61 & 19.90 & 21.01\\
135$^{\circ}$  & 20.86 & 20.21 & 20.08 & 20.01 & 19.61 & 20.12\\
180$^{\circ}$  & 19.30 & 18.73 & 18.78 & 18.97 & 18.92 & 18.80\\
\hline 
\hline 
\end{tabular} 
\begin{tablenotes}
     \item[1] Note: The sky spectra is calculated by \emph{SkyCalc} when $airmass=1.0$, PWV = 2.5 mm. We calculated the sky background brightness $m^{AB}$ when $airmass=1.2$.
   \end{tablenotes}
\end{table}

\subsection{Limiting Magnitudes of WFST}
Assuming the signal to noise ratio of WFST in all bands for a point source is $S/N$, we can write the formula of the $S/N$ as :
\begin{equation}
    \begin{aligned}
        \frac{S}{N}
        %= & \frac{N_{source}}{\sqrt{N_{source}+N_{sky}+N_{dark}+N_{readout}}}\\
        = & \frac{S\cdot A\cdot \tau}{\sqrt{S\cdot A\cdot \tau + 2\cdot n_{pix}\cdot [(Sky\cdot A\cdot \alpha_{pix} +D)\cdot \tau +R^2]}}
    \end{aligned}
    \label{SNR}
\end{equation}
where $S$ is the source signal with a constant spectral flux, $\tau$ is the standard exposure time (30 s), $A$ is the effective area of the primary mirror ($\rm \sim 4.12\times10^{4}\; cm^2$), $\alpha_{pix}=0.111$ arcsec$^2$ is the area of one pixel, $D$ is the dark current of the CCD ($D$ $\rm =0.005\; e^{-}/pixel/s$, @$\rm -100^{\circ} C$), $R^2$ is the readout noise of the CCD ($R$ $\rm =8\; e^{-}\; rms$), $n_{pix}$ is the total pixel number in the point spread function (PSF),
the usage of a factor 2 is because we assume the calculation is performed on sky subtracted images.
An optimal PSF aperture of 1.18 times of the full width at half maximum ($F\!W\!H\!M$) is adopted for a non-Adaptive Optics case according to the \emph{Integration Time Calculator (ITC)} of \emph{Gemini}$\footnote{\url{https://www.gemini.edu/observing/resources/itc/itc-help}}$. %\label{footnote2}
And the $F\!W\!H\!M$ of the seeing degrades with the $airmass$ and the wavelength as $(airmass)^{0.6}\times \lambda_{eff}^{-0.2}$.
Here $\lambda_{eff}$ takes the value of 356.17, 476.34, 620.57, 753.07, 870.45, 612.15 $\rm nm$ in the six bands $ugrizw$ given by the Equation \ref{lambdaeff}.
\begin{equation}
    \lambda_{eff}=\frac{\int_{0}^{\infty} \lambda T_{band}\;d\lambda}{\int_{0}^{\infty} T_{band}\;d\lambda}
    \label{lambdaeff}
\end{equation}
With the $seeing=0.75$ arcsec measured by \citet{Deng+etal+2021} at 500 $\rm nm$ \citep{Tokovinin+2003+etal}, we estimated the seeing values in different bands and airmass conditions. 

The sky signal actually lands on the detector is:
\begin{equation}
    Sky=\int_{0}^{\infty} f_{\lambda}T_{opt}T_{band}QE_{CCD}\;d\lambda
\end{equation}
where $T_{opt}$ is the throughput of the optics (including the primary mirror, ADC and the 5 corrector lenses),  $QE_{CCD}$ is the quantum efficiency of the CCD.

We can solve the Equation \ref{SNR} to obtain the signal of an astronomical object required at the detection limit of $S/N$ = 5 and find the corresponding limiting magnitude $m_{lim}$:

\begin{equation}
    m_{lim}=-2.5\times \log_{10}\left(\frac{S}{0.61\cdot ZP_{lim}}\right)
\end{equation}
A factor 0.61 is used because according to the description of \emph{ITC}, the 1.18 $F\!W\!H\!M$
sized aperture will contain 61\% energy of a point source. The $ZP_{lim}$ is the system zero point flux: 
\begin{equation}
    ZP_{lim}=\int_{0}^{\infty}flux_{AB}T_{atmo}T_{opt}T_{band}qe_{CCD}\,d\lambda
\end{equation}

Table \ref{tab1} lists the calculated limiting magnitudes of $ugrizw$ six bands.
We calculated the limiting magnitudes of WFST
at different Moon phases when the sky background brightness is V=22.0 mag and 22.3 mag, respectively. The results of a single exposure of 30 s and of coadded 100 frames with a total integration time of $100\times30$ s are listed. It shows that WFST can reach 23.42 (25.95) mag in the $g$ band with a 30 s ($100\times30$ s) exposure under the conditions with the sky background brightness V=22.3 mag, seeing =0.75 arcseconds, $airmass=1.2$ and PWV=2.5 mm. If the sky background is V=22.0 mag, the above values are 23.32 (25.85) mag for 30 s ($100\times30$ s).
\begin{table}[hbtp]
 \centering 
 \caption{$5\sigma$ limiting magnitudes of WFST when airmass=1.2, seeing = 0.75 arcsec, precipitable water vapour
(PWV) = 2.5 mm and Moon-object separation is $45^{\circ}$.} 
 \label{tab1} 
 \begin{tabular}{c c c c c c c c c c} 
 \hline 
 \hline 
  Exposure time & Moon Phase & V band sky & $u$ & $g$ & $r$ & $i$ & $z$ & $w$ \\
\hline
\hline 
%22.30  & 30 s  & No Moon & 22.98 & 23.75 & 23.21 & 22.60 & 21.29 & 23.85\\
30 s  & 0$^{\circ}$ & 22.30 & 22.31 & 23.42 & 22.95 & 22.43 & 21.50 & 23.61\\
30 s  & 45$^{\circ}$ & 22.10 & 22.27 & 23.30 & 22.89 & 22.40 & 21.49 & 23.54\\
30 s  & 90$^{\circ}$ & 21.29 & 22.04 & 22.86 & 22.62 & 22.26 & 21.43 & 23.23\\
30 s  & 135$^{\circ}$ & 20.28 & 21.64 & 22.34 & 22.21 & 21.99 & 21.31 & 22.79\\
30 s  & 180$^{\circ}$ & 18.90 & 20.97 & 21.62 & 21.58 & 21.49 & 21.00 & 22.13\\ \cline{2-9}
%22.30  & 3000 s  & No Moon & 25.88 & 26.38 & 25.80 & 25.18 & 23.84 & 26.38\\
$100\times30$ s  & 0$^{\circ}$ & 22.30 & 24.86 & 25.95 & 25.48 & 24.96 & 24.03 & 26.13\\
$100\times30$ s  & 45$^{\circ}$ & 22.10 & 24.82 & 25.84 & 25.42 & 24.93 & 24.02 & 26.06\\
$30\times100$ s  & 90$^{\circ}$ & 21.29 & 24.58 & 25.38 & 25.14 & 24.78 & 23.96 & 25.74\\
$100\times30$ s  & 135$^{\circ}$ & 20.28 & 24.17 & 24.85 & 24.72 & 24.51 & 23.83 & 25.30\\
$100\times30$ s  & 180$^{\circ}$ & 18.90 & 23.48 & 24.12 & 24.09 & 24.01 & 23.51 & 24.64\\
\hline 
%22.00  & 30 s  & No Moon & 22.90 & 23.63 & 23.08 & 22.47 & 21.15 & 23.70\\
30 s  & 0$^{\circ}$ & 22.00 & 22.26 & 23.32 & 22.83 & 22.30 & 21.37 & 23.47\\
30 s  & 45$^{\circ}$ & 21.80 & 22.21 & 23.19 & 22.77 & 22.28 & 21.37 & 23.40\\
30 s  & 90$^{\circ}$ & 20.99 & 21.95 & 22.74 & 22.48 & 22.12 & 21.29 & 23.09\\
30 s  & 135$^{\circ}$ & 19.98 & 21.52 & 22.19 & 22.07 & 21.85 & 21.18 & 22.64\\
30 s  & 180$^{\circ}$ & 18.60 & 20.83 & 21.47 & 21.44 & 21.35 & 20.86 & 21.99\\ \cline{2-9}
%22.00  & 3000 s  & No Moon & 25.73 & 26.23 & 25.65 & 25.04 & 23.69 & 26.23\\
$100\times30$ s  & 0$^{\circ}$ & 22.00 & 24.81 & 25.85 & 25.36 & 24.83 & 23.90 & 25.99\\
$100\times30$ s  & 45$^{\circ}$ & 21.80 & 24.76 & 25.72 & 25.30 & 24.80 & 23.89 & 25.92\\
$100\times30$ s  & 90$^{\circ}$ & 20.99 & 24.48 & 25.25 & 25.01 & 24.65 & 23.83 & 25.60\\
$100\times30$ s  & 135$^{\circ}$ & 19.98 & 24.05 & 24.71 & 24.58 & 24.37 & 23.70 & 25.15\\
$100\times30$ s  & 180$^{\circ}$ & 18.60 & 23.34 & 23.98 & 23.95 & 23.86 & 23.38 & 24.49\\
\hline 
\hline 
\end{tabular} 
    \begin{tablenotes}
     \item[1] Note: The V-band sky brightness is the Zenith sky background magnitudes.
   \end{tablenotes}
\end{table}

\section{Discussion and Conclusions}
\label{sect:conclusion}
\begin{figure}[h]  %[htbp]中的h是浮动的意思
    \centering    %居中
    \includegraphics[width=0.49\textwidth]{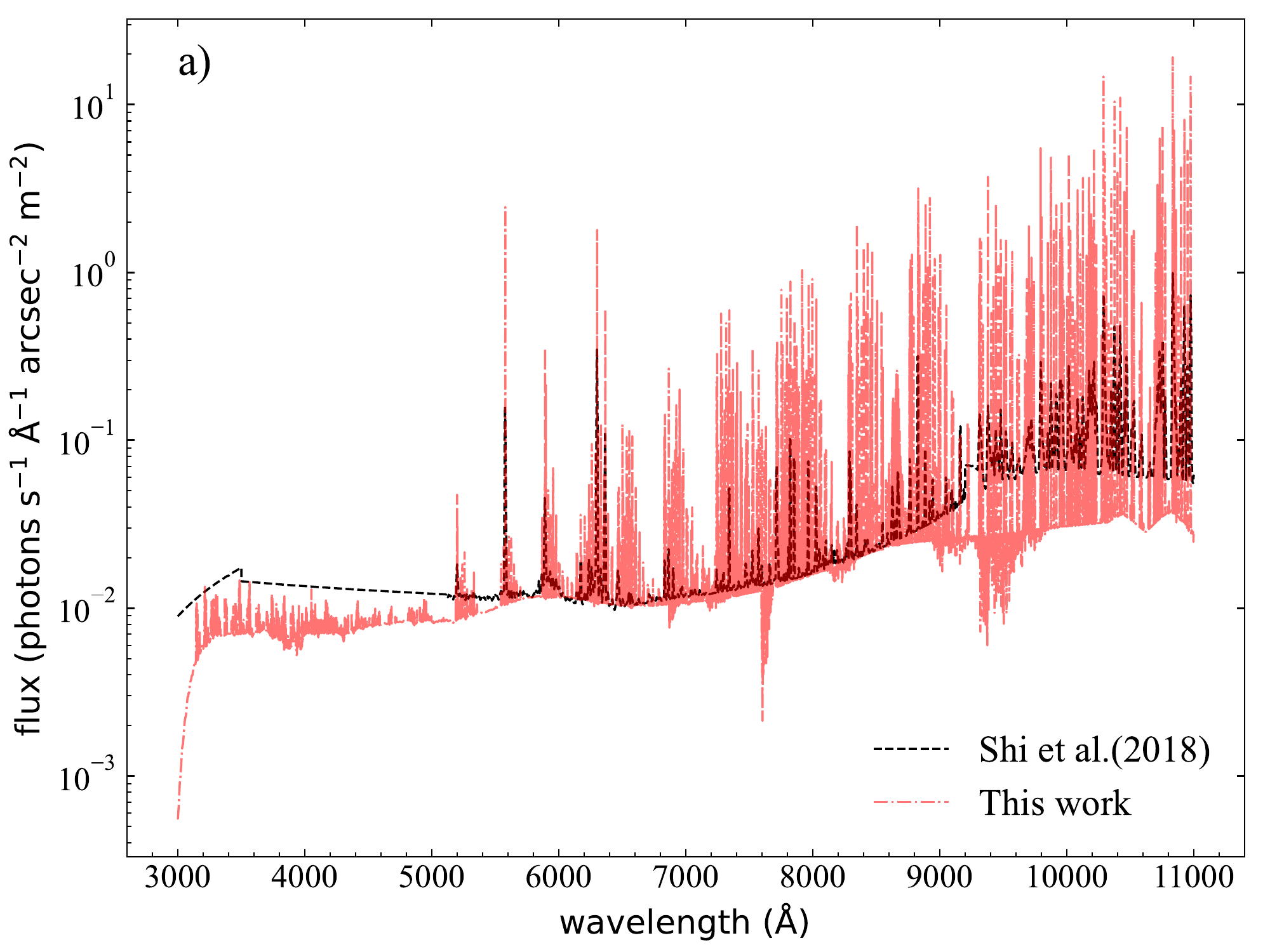}   %以行宽的0.5倍大小显示
    \includegraphics[width=0.49\textwidth]{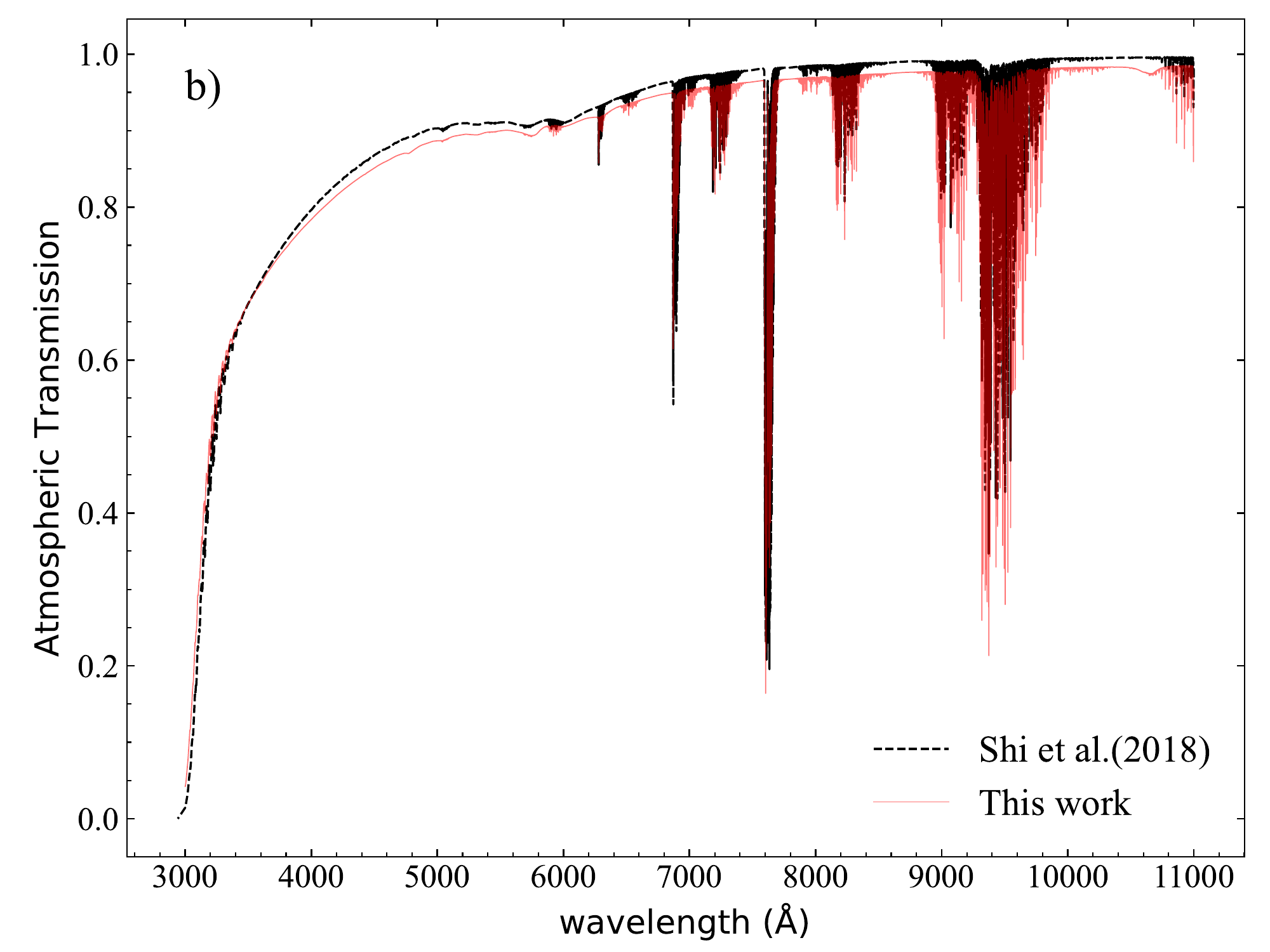}   %以行宽的0.5倍大小显示
    \caption{(a) The red line shows the sky background spectrum of Lenghu site at New Moon night (Moon phase $\theta=0^{\circ}$), $m_V=22.3$ $\rm mag$, $airmass = 1.2$. The black dashed line shows the Hawaii sky background spectrum scaled into $m_V=22.3$ $\rm mag$ and $airmass = 1.2$ in \citet{Shi+etal+2018}; (b) The red line shows the atmospheric transmittance curve of Lenghu site estimated by \emph{SkyCalc} in this work. The black dashed line shows the atmospheric transmittance curve of Shiquanhe astronomical site at an altitude of 5130 m at the Ali Area on the Tibetan Plateau estimated by the software \emph{MODTRAN}.} %  %大图名称
    \label{V.S.}  %图片引用标记
\end{figure}

%\subsection{Conclusions}
In the current work, by considering the observational conditions of WFST, including throughput, quantum efficiency, the noise, the area of the
primary mirror and the sky background brightness, we compute the limiting magnitudes of WFST. We get the sky background magnitudes in AB magnitude system in the Lenghu site at the New Moon night when $airmass=1.2$: $u,g,r,i,z,w$=23.27, 22.82, 21.80, 20.99, 20.05, 21.78 $\rm mag\, arcsec^{-2}$. For the Lenghu darkest night condition (V=22.3 $\rm mag\,arcsec^{-2}$) and a exposure time of 30 s, the $5\sigma$ limiting magnitudes of WFST are: 
$u_{lim},g_{lim},r_{lim},i_{lim},z_{lim},w_{lim}=$ 22.31, 23.42, 22.95, 22.41, 21.48, 23.60 $\rm mag$. 
The current estimates of limiting magnitudes are deeper than those in \citet{Shi+etal+2018}. This is because the current total throughput of WFST is higher than previous value, especially the throughput increases by $\sim50\%$ from $\sim0.4$ to $\sim0.6$ in $gri$ bands (see Figure \ref{Fig2}(a)), and the current Dark night sky background is lower than the previous estimation. Figure \ref{V.S.} compares the sky spectrum of New Moon night of the Lenghu site and the atmospheric transmittance curve between this work and \citet{Shi+etal+2018}. We used \emph{SkyCalc} to estimate the sky background spectrum and atmospheric transmittance, while \citet{Shi+etal+2018} used the software \emph{MODTRAN}$\footnote{\url{http://modtran.spectral.com/}}$ for estimating the atmospheric transmittance at the 5130 m Ali area and used a Hawaii sky background spectrum as a sky background spectral template. The Hawaii sky brightness in $ugz$ bands is brighter than the current model when we scaled both of them into the same conditions of $m_V=22.3$ $\rm mag\,arcsec^{-2}$ and $airmass = 1.2$ (see Figure \ref{V.S.} (a)), \citet{Shi+etal+2018} assumed the V band sky brightness is $m_V=21.50\, \rm mag\,arcsec^{-2}$. There is little difference between the current atmospheric transmittance model and the spectrum in \citet{Shi+etal+2018} (see Figure \ref{V.S.}(b)). Our scaled atmospheric transmittance is close to the model of \emph{MODTRAN}. %\label{footnote2}

\begin{figure}[h]  %[htbp]中的h是浮动的意思
    \centering    %居中
    \includegraphics[width=0.32\textwidth]{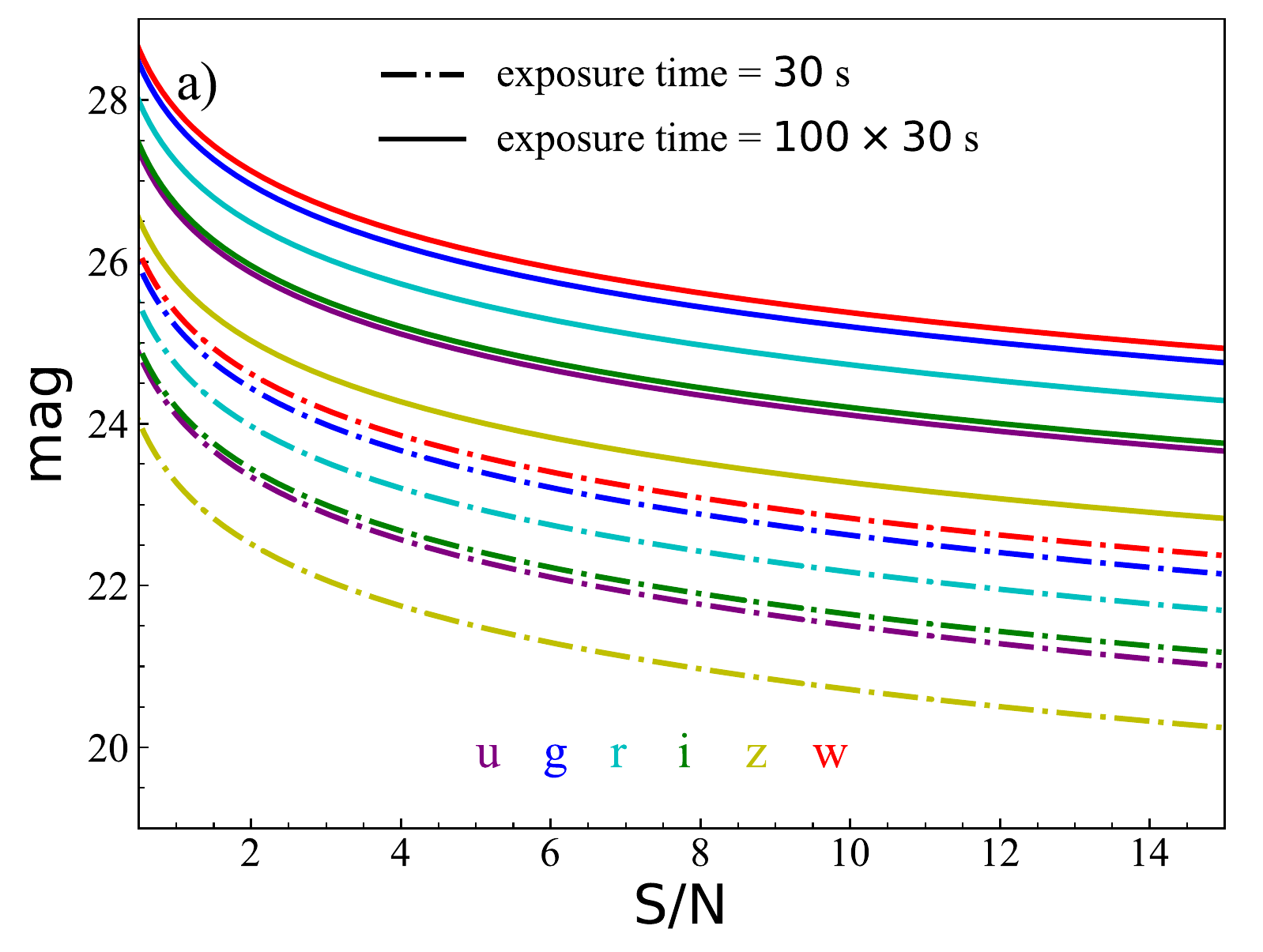}   %以行宽的0.5倍大小显示
    \includegraphics[width=0.32\textwidth]{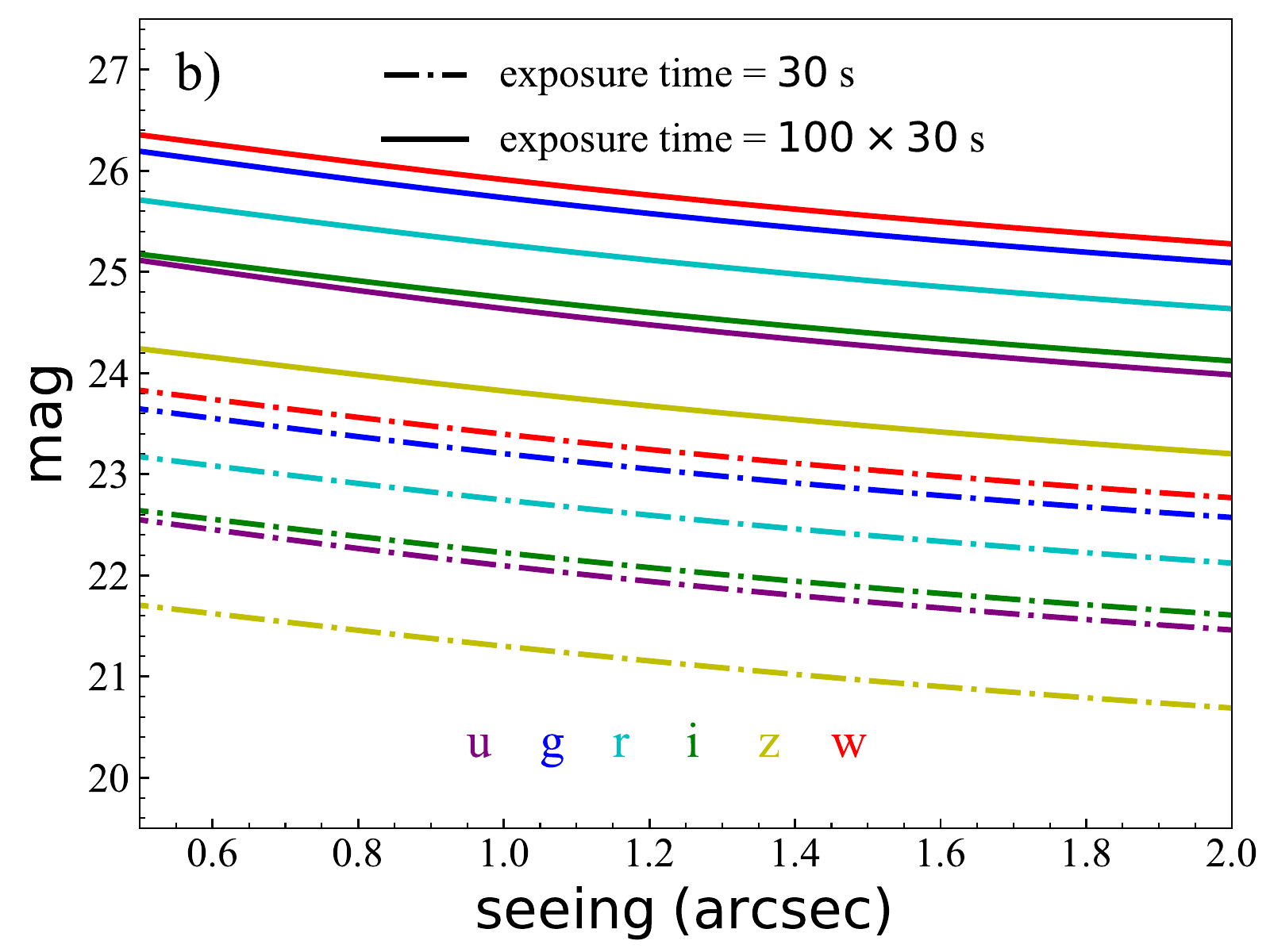}   %以行宽的0.5倍大小显示
    \includegraphics[width=0.32\textwidth]{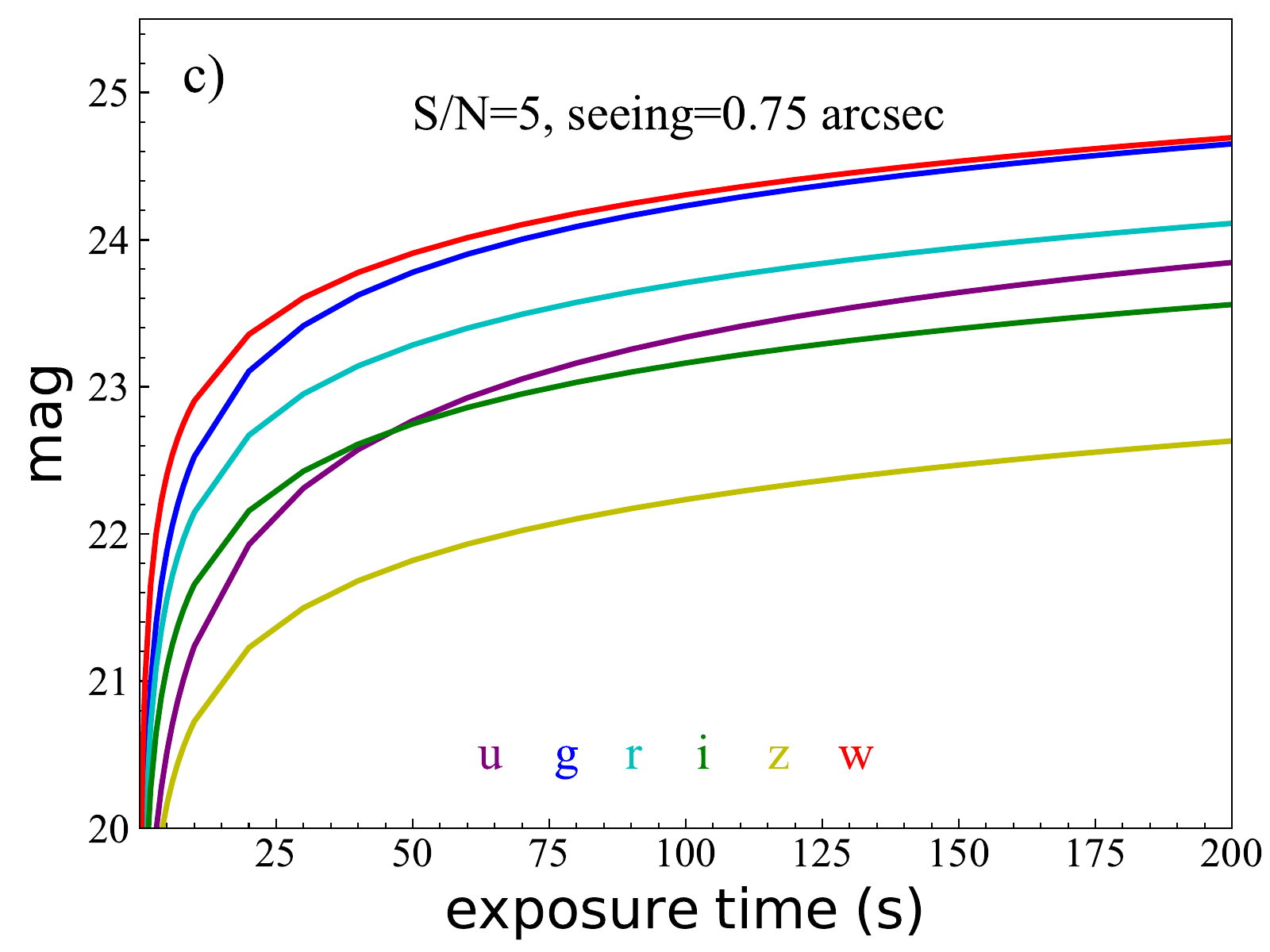}   %以行宽的0.5倍大小显示
    \caption{(a) The limiting magnitudes of different $S/N$ values when the exposure time is 30 s (dot-dashed line) and $100\times30$ s (solid line); (b) The $5\sigma$ limiting magnitudes of different seeing conditions when the exposure time is 30 s (dot-dashed line) and $100\times30$ s (solid line); (c) The $5\sigma$ limiting magnitudes of different exposure times when the seeing is 0.75 arcsec. The conditions for a New Moon night ($\theta=0^{\circ}$) and airmass = 1.2. Note: The limiting magnitude curves of the $g$ band (blue) and the $w$ band (red) are so close that we can not distinguish the two curves easily. } %  %大图名称
    \label{Fig5}  %图片引用标记
\end{figure}

%\subsection{Discussion}
We also obtain the limiting magnitudes of WFST under various conditions (Figure \ref{Fig5}).
In Figure \ref{Fig5}, the panel (a) shows the WFST limiting magnitudes of different signal-to-noise ratio when the exposure time equals to 30 s and $100\times30$ s respectively, the panel (b) shows the limiting magnitudes of different seeing conditions when signal-to-noise ratio = 5 and the exposure time = 30 s, $100\times30$ s respectively, and the panel (c) shows the $5\sigma$ limiting magnitudes of different exposure times. These results are calculated with the sky spectrum scaled into $airmass=1.2$ condition at a New Moon night (Moon phase $\theta=0^{\circ}$).

The WFST survey data will cover the entire northern sky. Its stacked scientific image data can be used to study asteroids, solar system, galaxies and cosmology. Its light curves can be used to discover variable objects. The estimated WFST limiting redshift of Type Ia supernovae (SNe~Ia) can reach z$\sim$0.64 (luminosity distance $\sim$ $6.3\times 10^3$ Mpc) and z$\sim$1.67 ($\sim1.2 \times 10^4$ Mpc) when the exposure time is 30 s and $100\times30$ s.
SNe~Ia can be used to constrain the dark energy in the universe \citep{Riess+etal+1998} and directly measure the Hubble constant \citep{Riess+etal+2022}. By simulating observations of the SNe~Ia with the WFST at the Lenghu site, \citet{Hu+etal+2022} estimate that above $10^{4}$ pre-maximum SNe~Ia will be discovered in one-year during the wide or deep observations, which suggests that WFST will be a powerful facility in revealing the physics of SNe~Ia.
\citet{Lin+etal+2022} computed the prospects of finding Tidal Disruption Events (TDEs) with the WFST. Their mock observations on 440 deg$^2$ field (CosmoDC2 catalogue) show that $\sim 30$ TDEs can be found per year if observed at $ugrizw$ bands with 30 s exposures every 10 days. According to \citet{Gao+etal+2022}, the event rate for galaxy-lensed orphan afterglows of $\gamma$-ray bursts (GRBs) is to be less than 0.7 yr$^{-1}$ for the whole sky survey of the WFST. \citet{Yu+etal+2021} estimated the multi-messenger detection rate of Binary Neutron Star Mergers is about 300-3500 yr$^{-1}$ with a GECAM-like detector for $\gamma$-ray emissions and an LSST/WFST detector for optical afterglows. \citet{Zhu+etal+2021a} and \citet{Zhu+etal+2022} showed that the optimal detection rates of the KN-dominated and AG-dominated GRB afterglows events are $\sim$0.2/0.5/0.8/20 yr$^{-1}$ and $\sim$500/300/600/3000 yr$^{-1}$ for ZTF/Mephisto/WFST/LSST, respectively. There are also some studies looking forward to detecting Active galactic nucleus (AGN) and researching AGN physics using WFST survey data (Xu-Fan Hu et al. in preparation; Su et al. in preparation). 

There are large sky survey telescopes that have been built around the world, and a number of large sky survey telescopes are being built. These projects have produced or will generate a large amount survey data and have an important impact in all fields of astronomy. Among them, the WFST will be completed in 2023. In the future, WFST (\citealt{Lin+etal+2022}; \citealt{Shi+etal+2018}), together with Mephisto (\citealt{Lei+etal+2020}; \citealt{Lei+etal+2022}; Chen et al. in preparation), Pan-STARRS (\citealt{Jedicke+etal+2007}; \citealt{Chambers+etal+2016}), SkyMapper (\citealt{Schmidt+etal+2005}; \citealt{Rakich+etal+2006}), ZTF (\citealt{Bellm+etal+2019}; \citealt{Graham+etal+2019}) and other telescopes will be able to carry out relay observations of the entire sky with large percentage time coverage, which will greatly enhance the development of the time-domain astronomy. 

%With the limiting magnitude, the detection depth can be calculated, so as the detection rate of various variable sources, such as supernova, kilonova, tidal disruption events (TDEs), active galaxy nuclears (AGNs), etc. \citet{Lin+etal+2022} computed the prospects of finding TDEs with the WFST. Their mock observations on 440 deg$^2$ field (CosmoDC2 catalogue) shows that $\sim 30$ TDEs can be found per year if observed at $ugrizw$ bands with 30-second exposure every 10 days. According to \citet{Gao+etal+2022}, the event rate for galaxy-lensed orphan afterglows of gamma-ray bursts (GRBs) is to be less than 0.7 yr$^{-1}$ for the whole sky survey of the WFST. \citet{Yu+etal+2021} estimated the multimessenger detection rate of Binary Neutron Star Mergers rate is about 300-3500 yr$^{-1}$ with a GECAM-like detector for $\gamma$-ray emissions and an LSST/WFST detector for optical afterglows. \citet{Zhu+etal+2021a} and \citet{Zhu+etal+2021b} showed that the optimal detection rates of the KN-dominated (AG-dominated) events are $\sim$0.2/0.5/0.8/20 yr $^{-1}$  ($\sim$500/300/600/3000 yr $^{-1}$) for ZTF/Mephisto/WFST/LSST, respectively. There are also some studies looking forward to detect AGN and research AGN physics using WFST survey data (\citealt{Lei+etal+2022}; X. F. Hu et al. 2022, in preparation; Z. B. Su et al. 2022, in preparation).

\begin{acknowledgements}
This work is supported by the Strategic Priority Research Program of Chinese Academy of Sciences (Grant No. XDB 41000000, XDB 41010105), the National Science Foundation of China (NSFC, Grant No. 12233008, 12173037, 11973038), the China Manned Space Project (No. CMS-CSST-2021-A07) and the Cyrus Chun Ying Tang Foundations.
We thank Fredrik T Rantakyrö and Rodolfo Angeloni from Gemini Observatory for their patient elaboration on the Hawaii sky spectrum model and sky brightness measurements.

\end{acknowledgements}

%\appendix                  %%appendicial material is supported

%\section{This shows the use of appendix}
%A postscript file is actually an ASCII text file (you may even edit it).
%However, you need to transfer a PDF file or any compressed or packaged
%file in binary mode when using FTP.

%\section{What is SCI?}
%SCI is the abbreviation of Science Citation Index system powered by
%the Institute for Scientific Information (ISI). For details please
%visit {\it http://apps.isiknowledge.com}.

\bibliographystyle{raa}
\bibliography{references}

\begin{thebibliography}{39}
\providecommand\natexlab[1]{#1}
\providecommand\JournalTitle[1]{#1}

\bibitem[{Bellm}(2014)]{Bellm+etal+2014}
{Bellm}, E. 2014, in The Third Hot-wiring the Transient Universe Workshop, ed.
  P.~R. {Wozniak}, M.~J. {Graham}, A.~A. {Mahabal}, \& R.~{Seaman}, 27

\bibitem[{Bellm} {et~al.}(2019)]{Bellm+etal+2019}
{Bellm}, E.~C., {Kulkarni}, S.~R., {Barlow}, T., {et~al.} 2019, \pasp, 131,
  068003

\bibitem[{Bessell} \& {Murphy}(2012)]{Bessell+etal+2012}
{Bessell}, M., \& {Murphy}, S. 2012, \pasp, 124, 140

\bibitem[{Bessell}(1990)]{Bessell+etal+1990}
{Bessell}, M.~S. 1990, \pasp, 102, 1181

\bibitem[{Chambers} \& {Pan-STARRS Team}(2016)]{Chambers+etal+2016}
{Chambers}, K.~C., \& {Pan-STARRS Team}. 2016, in American Astronomical Society
  Meeting Abstracts, Vol. 227, American Astronomical Society Meeting Abstracts
  \#227, 324.07

\bibitem[{Chen} {et~al.}(2019)]{Chen+etal+2019}
{Chen}, J., {Zhang}, H.-f., {Wang}, J., {Chen}, J.-t., \& {Zhang}, J. 2019, in
  Society of Photo-Optical Instrumentation Engineers (SPIE) Conference Series,
  Vol. 11101, Material Technologies and Applications to Optics, Structures,
  Components, and Sub-Systems IV, 111010D

\bibitem[{Deng} {et~al.}(2021)]{Deng+etal+2021}
{Deng}, L., {Yang}, F., {Chen}, X., {et~al.} 2021, \nat, 596, 353

\bibitem[{Fukugita} {et~al.}(1996)]{Fukugita+etal+1996}
{Fukugita}, M., {Ichikawa}, T., {Gunn}, J.~E., {et~al.} 1996, \aj, 111, 1748

\bibitem[{Gao} {et~al.}(2022)]{Gao+etal+2022}
{Gao}, H.-X., {Geng}, J.-J., {Hu}, L., {et~al.} 2022, \mnras

\bibitem[{Graham} {et~al.}(2019)]{Graham+etal+2019}
{Graham}, M.~J., {Kulkarni}, S.~R., {Bellm}, E.~C., {et~al.} 2019, \pasp, 131,
  078001

\bibitem[{Hlozek} {et~al.}(2019)]{Hlozek+etal+2019}
{Hlozek}, R., {Albert}, J., {Balogh}, M., {et~al.} 2019, in Canadian Long Range
  Plan for Astronomy and Astrophysics White Papers, Vol. 2020, 51

\bibitem[{Hu} {et~al.}(2022)]{Hu+etal+2022}
{Hu}, M., {Hu}, L., {Jiang}, J.-a., {et~al.} 2022, Universe, 9, 7

\bibitem[{Jedicke} \& {Pan-STARRS}(2007)]{Jedicke+etal+2007}
{Jedicke}, R., \& {Pan-STARRS}. 2007, in AAS/Division for Planetary Sciences
  Meeting Abstracts, Vol.~39, AAS/Division for Planetary Sciences Meeting
  Abstracts \#39, 8.02

\bibitem[{Jones} {et~al.}(2013)]{Jones+etal+2013}
{Jones}, A., {Noll}, S., {Kausch}, W., {Szyszka}, C., \& {Kimeswenger}, S.
  2013, \aap, 560, A91

\bibitem[{Kent}(1994)]{Kent+etal+1994}
{Kent}, S.~M. 1994, \apss, 217, 27

\bibitem[{Krisciunas} \& {Schaefer}(1991)]{Krisciunas+etal+1991}
{Krisciunas}, K., \& {Schaefer}, B.~E. 1991, \pasp, 103, 1033

\bibitem[{Lei} {et~al.}(2022)]{Lei+etal+2022}
{Lei}, L., {Chen}, B.-Q., {Li}, J.-D., {et~al.} 2022, Research in Astronomy and
  Astrophysics, 22, 025004

\bibitem[{Lei} {et~al.}(2021)]{Lei+etal+2020}
{Lei}, L., {Li}, J.~D., {Wu}, J.~T., {Jiang}, S.~Y., \& {Chen}, B.~Q. 2021,
  Astronomical Research \& Technology, 18, 115121

\bibitem[{Lin} {et~al.}(2022)]{Lin+etal+2022}
{Lin}, Z., {Jiang}, N., \& {Kong}, X. 2022, \mnras, 513, 2422

\bibitem[{Liu}(2019)]{Liu+etal+2019}
{Liu}, X. 2019, in Galactic Archaeology in the Gaia Era, 14

\bibitem[{Lou} {et~al.}(2020)]{Lou+etal+2020}
{Lou}, Z., {Liang}, M., {Zheng}, X.~Z., {et~al.} 2020, in Society of
  Photo-Optical Instrumentation Engineers (SPIE) Conference Series, Vol. 11445,
  Society of Photo-Optical Instrumentation Engineers (SPIE) Conference Series,
  114454A

\bibitem[{Lou} {et~al.}(2016)]{Lou+etal+2016}
{Lou}, Z., {Liang}, M., {Yao}, D., {et~al.} 2016, in Society of Photo-Optical
  Instrumentation Engineers (SPIE) Conference Series, Vol. 10154, Society of
  Photo-Optical Instrumentation Engineers (SPIE) Conference Series, 101542A

\bibitem[{Moehler} {et~al.}(2014)]{Moehler+etal+2014}
{Moehler}, S., {Modigliani}, A., {Freudling}, W., {et~al.} 2014, \aap, 568, A9

\bibitem[{Noll} {et~al.}(2012)]{Noll+etal+2012}
{Noll}, S., {Kausch}, W., {Barden}, M., {et~al.} 2012, \aap, 543, A92

\bibitem[{Petrova} {et~al.}(2021)]{Petrova+2021+etal}
{Petrova}, E., {Podladchikova}, T., {Veronig}, A.~M., {et~al.} 2021, \apjs,
  254, 9

\bibitem[{Rakich} {et~al.}(2006)]{Rakich+etal+2006}
{Rakich}, A., {Blundell}, M., {Pentland}, G., {et~al.} 2006, in Society of
  Photo-Optical Instrumentation Engineers (SPIE) Conference Series, Vol. 6267,
  Society of Photo-Optical Instrumentation Engineers (SPIE) Conference Series,
  ed. L.~M. {Stepp}, 62670E

\bibitem[{Riess} {et~al.}(1998)]{Riess+etal+1998}
{Riess}, A.~G., {Filippenko}, A.~V., {Challis}, P., {et~al.} 1998, \aj, 116,
  1009

\bibitem[{Riess} {et~al.}(2022)]{Riess+etal+2022}
{Riess}, A.~G., {Yuan}, W., {Macri}, L.~M., {et~al.} 2022, \apjl, 934, L7

\bibitem[{Schmidt} {et~al.}(2005)]{Schmidt+etal+2005}
{Schmidt}, B.~P., {Keller}, S.~C., {Francis}, P.~J., \& {Bessell}, M.~S. 2005,
  in American Astronomical Society Meeting Abstracts, Vol. 206, American
  Astronomical Society Meeting Abstracts \#206, 15.09

\bibitem[{Shi} {et~al.}(2018)]{Shi+etal+2018}
{Shi}, D.~D., {Zheng}, X.~Z., {Zhao}, H.~B., {et~al.} 2018, Acta Astronomica
  Sinica, 59, 22

\bibitem[{Sparavigna}(2008)]{Sparavigna+2008+etal}
{Sparavigna}, A. 2008, arXiv e-prints, arXiv:0804.1941

\bibitem[{Tokovinin} {et~al.}(2003)]{Tokovinin+2003+etal}
{Tokovinin}, A., {Baumont}, S., \& {Vasquez}, J. 2003, \mnras, 340, 52

\bibitem[{Wang} {et~al.}(2016)]{Wang+etal+2016}
{Wang}, H., {Lou}, Z., {Qian}, Y., {Zheng}, X., \& {Zuo}, Y. 2016, Optical
  Engineering, 55, 035105

\bibitem[{Yu} {et~al.}(2021)]{Yu+etal+2021}
{Yu}, J., {Song}, H., {Ai}, S., {et~al.} 2021, \apj, 916, 54

\bibitem[{Yuan} {et~al.}(2021)]{Yuan+etal+2021}
{Yuan}, H.-B., {Deng}, D.-S., \& {Sun}, Y. 2021, Research in Astronomy and
  Astrophysics, 21, 074

\bibitem[{Yuan} {et~al.}(2020)]{Yuan+etal+2020}
{Yuan}, X., {Li}, Z., {Liu}, X., {et~al.} 2020, in Society of Photo-Optical
  Instrumentation Engineers (SPIE) Conference Series, Vol. 11445, Society of
  Photo-Optical Instrumentation Engineers (SPIE) Conference Series, 114457M

\bibitem[{Zhao} {et~al.}(2016)]{Zhao+etal+2016}
{Zhao}, H., {Li}, Y., \& {Zhang}, C. 2016, IEEE Geoscience and Remote Sensing
  Letters, 13, 1139

\bibitem[{Zhu} {et~al.}(2021)]{Zhu+etal+2021a}
{Zhu}, J.-P., {Wu}, S., {Yang}, Y.-P., {et~al.} 2021, arXiv e-prints,
  arXiv:2110.10469

\bibitem[{Zhu} {et~al.}(2022)]{Zhu+etal+2022}
{Zhu}, J.-P., {Yang}, Y.-P., {Zhang}, B., {Gao}, H., \& {Yu}, Y.-W. 2022, \apj,
  938, 147

\end{thebibliography}
\label{lastpage}
\end{CJK*}
\end{document}